\documentclass[aps,prx,bibliography,superscriptaddress,,twocolumn]{revtex4-2}
\usepackage[utf8]{inputenc}
\usepackage{amsmath}
\usepackage{physics}
\usepackage{mathtools}
\usepackage{amsfonts}
\usepackage{amssymb}
\usepackage{braket}
\usepackage{graphicx}
\usepackage{subfigure}
\usepackage{textcomp}
\usepackage{color}
\usepackage[dvipsnames]{xcolor}
\usepackage{mathrsfs}
\usepackage{natbib}
\usepackage{makecell}
\usepackage{soul}

\begin{document}
\title{Emergence of tunable exceptional points in altermagnet-ferromagnet junctions}
\author{Md Afsar Reja}
\email{afsarmd@iisc.ac.in}
\affiliation{Solid State and Structural Chemistry Unit, Indian Institute of Science, Bangalore 560012, India}
\author{Awadhesh Narayan}
\email{awadhesh@iisc.ac.in}
\affiliation{Solid State and Structural Chemistry Unit, Indian Institute of Science, Bangalore 560012, India}
\date{\today}

\begin{abstract}
The existence of exceptional points (EPs) -- where both eigenvalues and eigenvectors converge -- is a key characteristic of non-Hermitian physics. A newly-discovered class of magnets -- termed as altermagnets (AMs) -- are characterized by a net zero magnetization as well as spin-split bands. In this study, we propose the emergence of non-Hermitian physics at AM-ferromagnet (FM) junctions. We discover that such a junction hosts tunable EPs. We demonstrate that the positions of these emergent EPs can be tuned using an external applied magnetic field and show that for a critical value of the applied magnetic field the EPs can annihilate. Notably, the number and position of the EPs crucially depends on the type of AM and its orientation with respect to the FM. Our work puts forth a promising platform of exploration of non-Hermitian physics in an emerging class of magnetic materials. 
\end{abstract}

\maketitle

\section{Introduction}

Non-Hermitian (NH) systems have gained extensive attention in both theoretical and experimental research in recent years~\cite{moiseyev2011non,bender2007making,ashida2020non,bergholtz2021exceptional,kawabata2019symmetry,banerjee2023non}. This growing interest stems from the realization of intriguing NH phenomena, such as EPs, the NH skin effect, exotic topological phases, generalized Bloch band theory, and extended symmetry classes, to name just a few. EPs, where both eigenvalues and eigenstates coalesce, are fundamentally unique to NH physics~\cite{heiss2012physics,kato2013perturbation}. EPs are associated with a plethora of fascinating phenomena, both in theory and applications, including NH topological phases linked with winding of eigenvalues and eigenvectors~\cite{ding2022non,banerjee2023tropical,mandal2021symmetry}, enhanced sensing~\cite{hodaei2017enhanced}, optical microcavities~\cite{chen2017exceptional}, and directional lasing technologies~\cite{peng2016chiral}. EPs have been realized across diverse experimental platforms, encompassing fields such as optics, photonics, electric circuits, and acoustics~\cite{parto2020non,miri2019exceptional,stehmann2004observation,choi2018observation,zhu2018simultaneous,shi2016accessing,zhu2018simultaneous}. In addition, they have been theoretically investigated at material interfaces, including topological insulator-FM~\cite{bergholtz2019non} and superconductor-FM\cite{cayao2022exceptional} junctions.

\begin{figure}[b]
\centering
\includegraphics[width=0.32\textwidth]{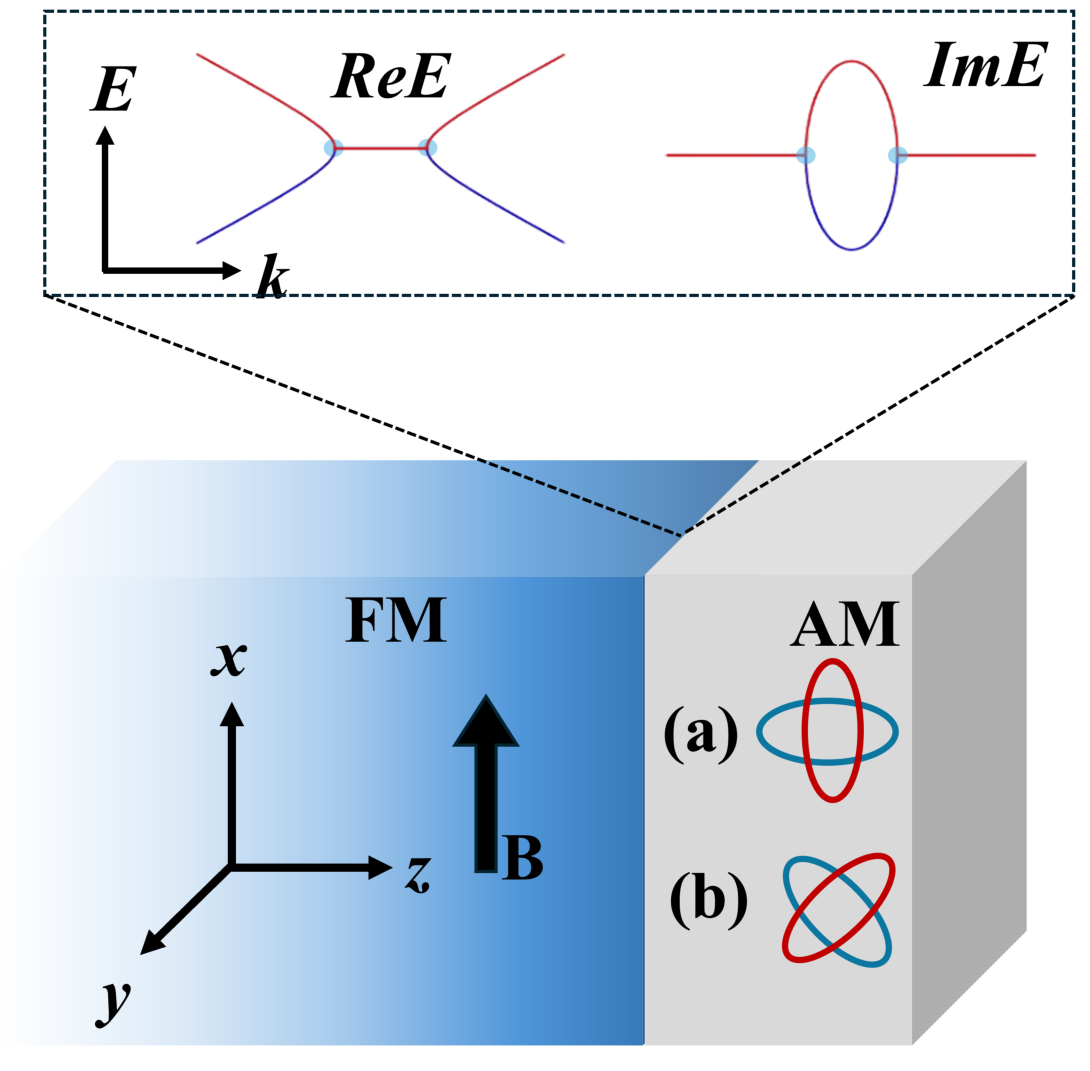}
\caption{\textbf{Proposed setup with AM-FM junction.} Illustration of the AM-FM junction at $z=0$, with FM region for $z<0$. Panels (a) and (b) depict the orientations of the Fermi surface of the $d$-wave AM. We propose tunable EPs at such junctions, as shown schematically. A magnetic field allows further control over the position of the EPs and their number.}
\label{Fig_junction}
\end{figure}

On the other hand, AMs are a newly-identified magnetic class distinct from conventional FMs and antiferromagnets~\cite{vsmejkal2022emerging,vsmejkal2022beyond}. In AMs, opposite spin sublattices are linked by a rotational symmetry rather than the conventional translational or inversion operations~\cite{vsmejkal2022beyond,litvin1974spin,brinkman1966theory}, which results in a subtle interplay of characteristics of both FMs and antiferromagnets. They demonstrate spin-split band structures, similar to FMs, but have zero total magnetization akin to antiferromagnets~\cite{vsmejkal2022beyond,vsmejkal2020crystal}. The library of altermagnetic materials has been expanding rapidly~\cite{vsmejkal2022beyond,fedchenko2024observation,vsmejkal2022emerging,yuan2020giant,vsmejkal2022emerging,vsmejkal2022beyond,krempasky2024altermagnetic,zhu2024observation,devaraj2024interplay}. Furthermore, AM junctions have been recently investigated for spin pumping~\cite{sun2023spin} and intriguing transport properties~\cite{das2023transport,lyu2024orientation,papaj2023andreev,ouassou2023dc,beenakker2023phase,niu2024electrically,giil2024superconductor,cheng2024field,das2024crossed,sukhachov2024thermoelectric,niu2024orientation,lu2024varphi,banerjee2024altermagnetic,nagae2024spin, zhang2024finite}.

In this study, we propose the emergence of non-Hermitian physics at AM-FM junctions. Through analytical and numerical calculations, we reveal that these junctions support tunable EPs. We demonstrate the presence of EPs in $d$-wave AM-FM junctions by observing the coalescence of eigenvalues and eigenvectors, and characterize them by examining the vorticity and the scaling of phase rigidity. We find that the positions of these EPs can be controlled using an external magnetic field, which can also be used to annihilate them. Significantly, the number and locations of the EPs depend on the orientation with respect to the FM lead and the type of AM, which we exemplify by means of $g$-wave and $i$-wave AM-FM junctions. Our work introduces a promising platform for exploring non-Hermitian physics in this new class of magnetic materials.

\begin{figure}
\centering
\includegraphics[width=0.48\textwidth]{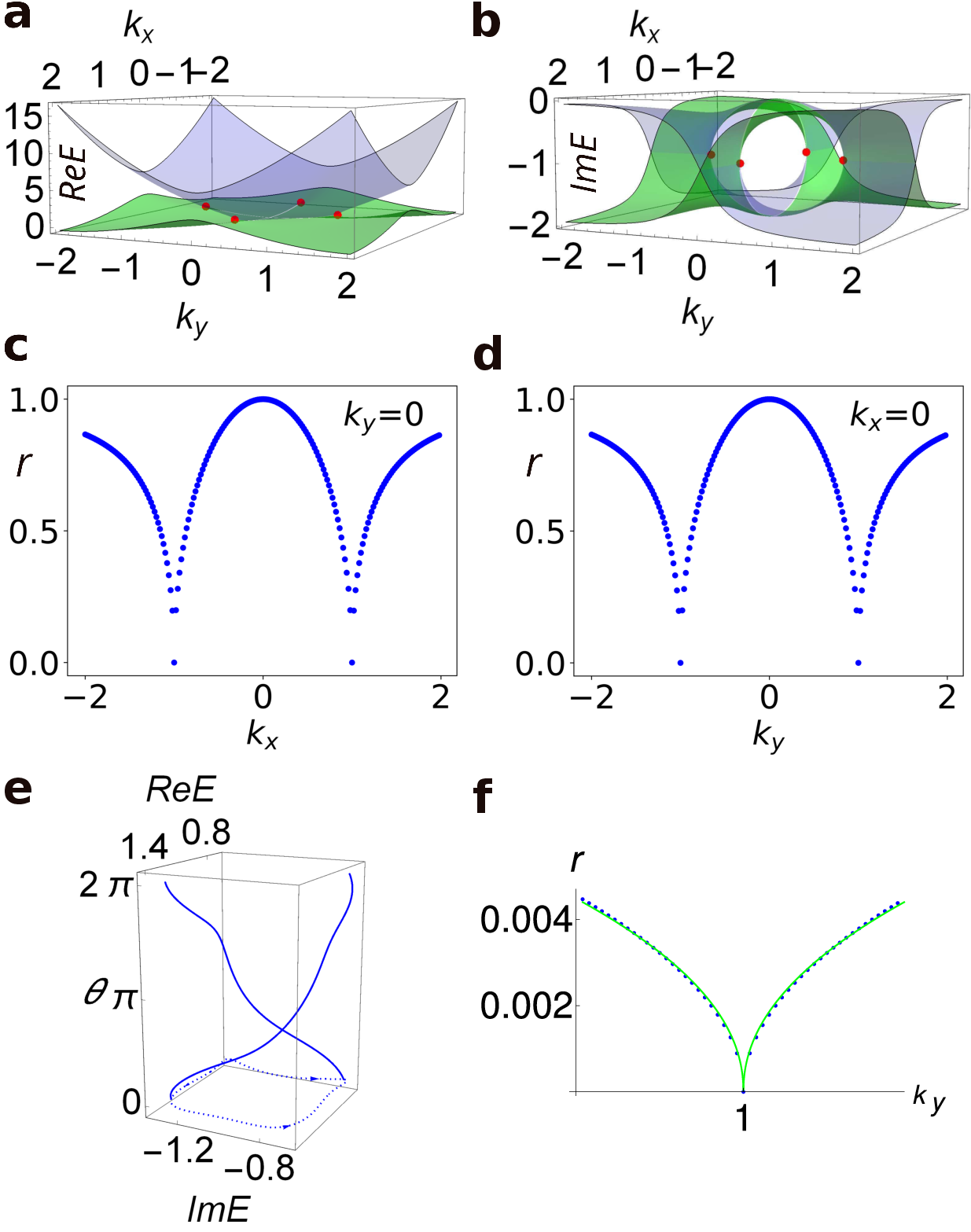}
\caption{\textbf{Emergent EPs in $d$ wave AM-FM junctions.} (a) Real and (b) imaginary parts of the eigenvalues. At the points marked by red circles the eigenvalues merge. The phase rigidity, $r$, along (c) $k_x$ and (d) $k_y$, respectively. Note that $r$ goes to zero at the marked points, signaling their exceptional nature. (e) Vorticity around one of the EPs. The interchange of the bands confirms the emergence of the EP. (f) Variation of phase rigidity near the EP. The blue dots are the calculated values of $r$, while the solid green curve is the fitted function, yielding an exponent of $\sim$0.47. This value is close to the exponent expected for a second-order EP. Here we choose $t=1$, $t_{j1}=1$, $\lambda=1$ and $\gamma=1$.}
\label{Fig_d_j1_B0}
\end{figure}

\section{Setup of altermagnet-ferromagnet junctions}

We consider an AM-FM junction by coupling a two-dimensional AM with a semi-infinite FM lead as depicted in Fig.~\ref{Fig_junction}. Region $z<0$ is the FM lead and $z=0$ is the interface. We treat the AM-FM junction as an open quantum system and model it with an effective Hamiltonian of the form,

\begin{equation}
H_\text{NH} = H_\text{AM} + \Sigma_L,
\label{eq_H_NH}
\end{equation}

where $H_\text{AM}$ denotes the Hamiltonian of the considered AM system (discussed later), and $\Sigma_L$ represents the self-energy arising from the semi-infinite FM lead. In the wide-band limit, the self-energy term becomes independent of both momentum and frequency, and can be analytically calculated as~\cite{ryndyk2009green,datta1997electronic,bergholtz2019non,cayao2022exceptional},

\begin{equation}
\Sigma_L=-i\Gamma\sigma_0 -i\gamma\sigma_z,
\end{equation}

where $\Gamma = \frac{\Gamma_{+} + \Gamma_{-}}{2}$, $\gamma = \frac{\Gamma_{+} - \Gamma_{-}}{2}$, and $\Gamma_{\pm} = \pi |t'|^2\rho_{\pm}^L$. Here $\rho_{\pm}^L = \frac{1}{t'\pi} \sqrt{1 - (\frac{\mu_L \pm m}{2t_z})^2}$ represents the surface densities of the lead for spin-up and spin-down channels and $t'$ is the hopping amplitude from the lead to the considered AM system. Here $\sigma_x, \sigma_y, \sigma_z$ are the Pauli matrices, $\sigma_0$ is the identity matrix, $t_z$ is hopping within the lead along the $z$ direction, $\mu_L$ is the chemical potential, and $m$ is intrinsic magnetization of the FM lead. The effective Hamiltonian of the junction becomes NH due to the imaginary contribution coming from the self-energy term. We will next analyze its effects on the exceptional properties of the system.

\begin{figure*}[t]
\centering
\includegraphics[width=0.85\textwidth]{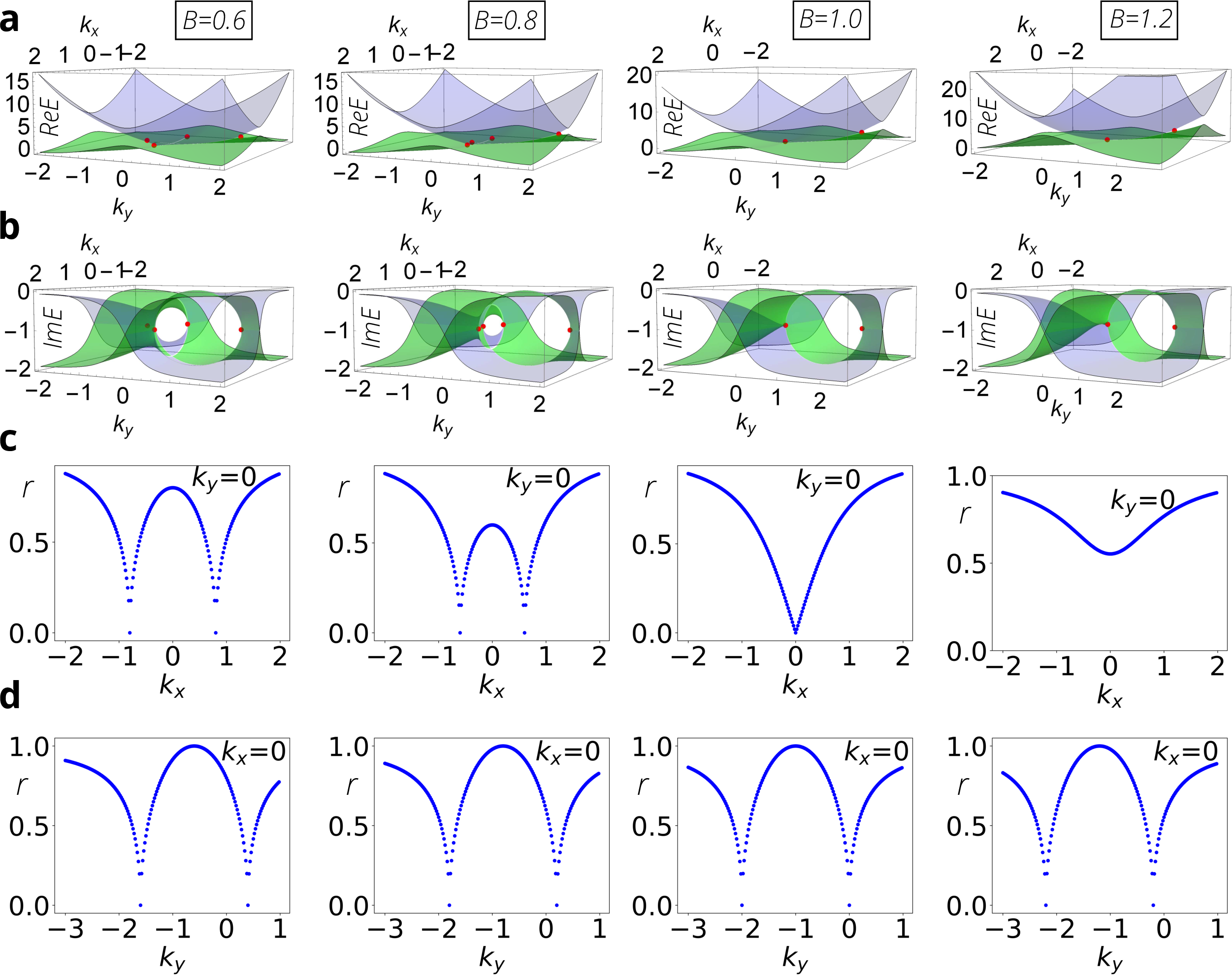}
\caption{\textbf{Magnetic field tuning of EPs.} Rows (a) and (b) show the real and imaginary parts of the energy eigenvalues with increasing magnetic field $B$ (left to right). We note that two of the EPs (denoted by red circles) move in opposite directions towards the origin along $k_x$. They merge at $B=1$. The other two EPs move linearly along the negative $k_y$ direction. Rows (c) and (d) present the phase rigidity, $r$, along $k_x$ and $k_y$, respectively, with increasing $B$. Note that the $r=0$ points come closer to the origin along $k_x$ and disappear above $B=1$. Along $k_y$, the $r=0$ points move along the negative $k_y$ direction. Here we set $t=1$, $t_{j1}=1$, $\lambda=1$ and $\gamma=1$.}
\label{Fig_d_B1}
\end{figure*}

\section{Exceptional points in d-wave altermagnets}

Let us consider the setup with an FM lead attached to a $d$-wave AM, whose Fermi surface is oriented as shown in Fig.~\ref{Fig_junction}(a). The Hamiltonian for a $d$-wave AM is given by~\cite{vsmejkal2022beyond}

\begin{equation}
\begin{aligned}
    H_\text{AM}^d=t(k_x^2+k_y^2)\sigma_0+ 2t_{j1}k_xk_y\sigma_z +\lambda(k_y\sigma_x-k_x\sigma_y),
\end{aligned}
\label{eq_d_j1}
\end{equation}

where $t$ is a hopping term, $t_{j1}$ is an AM-specific spin-dependent hopping, and $\lambda$ is the strength of the Rashba term. The effective NH Hamiltonian then becomes

\begin{equation}
 \begin{aligned}
        \Tilde{H} =& H_\text{AM}^d + \Sigma_L\\
              &= t(k_x^2 + k_y^2) \sigma_0 + 2t_{j1} k_x k_y \sigma_z\\ &+\lambda(k_y \sigma_x - k_x \sigma_y) + \Sigma_L.   
\end{aligned}
\label{eq_d_ham}
\end{equation}

The above equation can be written in the form $\Tilde{H}= \epsilon_0 + \mathbf{d} \cdot \mathbf{\sigma}$, with $\epsilon_0 \in \mathbb{C}$ and $\mathbf{d} = \mathbf{d}_R + i\mathbf{d}_I$ with $\mathbf{d}_R, \mathbf{d}_I \in \mathbb{R}^3$. For our model, $\mathbf{d}_R=(\lambda k_y, -\lambda k_x ,2t_{j1}k_xk_y)$ and $\mathbf{d}_I=(0,0,-\gamma)$. Now, the eigenvalues are given by $E_{\pm} = \epsilon_0 \pm \sqrt{\mathbf{d}_R^2 - \mathbf{d}_I^2 + 2i\mathbf{d}_R \cdot \mathbf{d}_I}$ and the conditions for the occurrence of EPs are $\mathbf{d}_R^2 = \mathbf{d}_I^2$ and $\mathbf{d}_R \cdot \mathbf{d}_I = 0$. For our AM-FM junction, we obtain the conditions

\begin{equation}
    \gamma^2=\lambda^2(k_x^2+k_y^2)+4t_{j1}k_x^2k_y^2, \quad \textcolor{black}{t_{j1}}\gamma k_xk_y=0.
 \label{eq_dj1_ep}
\end{equation}

Here $\gamma=0$ is a trivial solution and $\gamma$ must be non-zero to obtain EPs. Notably, we find that four EPs emerge at $(0,\pm \frac{\gamma}{\lambda})$, $(\pm \frac{\gamma}{\lambda},0)$ in the $k_x-k_y$ plane, when the conditions in Eq.~\ref{eq_dj1_ep} are simultaneously satisfied. 

For the remainder of the paper, for simplicity, we will set $t=1$, $t_{j1}=1$, $\gamma=1$, and $\lambda=1$. The real and imaginary parts of the eigenvalues are plotted in Fig.~\ref{Fig_d_j1_B0} (a) and (b), respectively. The red dots indicate the merging of the eigenvalues at $(0,\pm1), (\pm1,0)$. We further confirm the coalescing of the eigenvectors at the same points by calculating the phase rigidity, $r = \frac{\langle \Psi_L | \Psi_R\rangle}{ \langle \Psi_R | \Psi_R \rangle}$~\cite{heiss2012physics}. Here $ \Psi_L$ and $ \Psi_R$ denote the left and right eigenvectors respectively. Due to bi-orthogonalization, $r\rightarrow 0$ near EPs and approaches unity away from them. In Fig.~\ref{Fig_d_j1_B0}(c) and (d), we show $r$ along $k_x$ and $k_y$, respectively. We note that $r$ vanishes at $(0,\pm1)$ and $(\pm1,0)$, which confirms that not only the eigenvalues merge but also the eigenstates coalesce. 

As another characterization of the emergence of EPs, we calculate the vorticity, $\nu_{mn}$, given by~\cite{shen2018topological} 

\begin{equation}
  \nu_{mn}(C) = - \frac{1}{2\pi} \oint_C\nabla_\mathbf{k} \arg \left[ E_m(\mathbf{k}) - E_n(\mathbf{k}) \right] \cdot d\mathbf{k}.
\label{eq_vorticity}
\end{equation}

Here $E_m(\mathbf{k})$ and $E_n(\mathbf{k})$ denote the energy eigenvalues in the complex-energy plane and $C$ is a closed loop in the momentum space. In Fig.~\ref{Fig_d_j1_B0}(e), we plot the vorticity of one of the EPs [located at $(0,+1)$] by choosing $C$ around it. We clearly observe the braiding of bands, a direct signature of the underlying EP. This further confirms the appearance of EPs. To examine the order of EPs, we plot the phase rigidity in Fig.~\ref{Fig_d_j1_B0}(f). The blue dots indicate the calculated values of $r$, while the solid green line is the fitted function. The fitting revealed that the exponent of scaling of $r$ is $\sim$0.47, which is close to that expected for a second-order EP~\cite{bulgakov2006phase,rotter2009non}. \textcolor{black}{If we take into account the full tight binding model, the earlier results still hold true (for details see Appendix A).}
Having seen the emergence of EPs in AM-FM junctions, next, we will show how applying a magnetic field allows us to tune and control them.

\begin{figure}
\centering
\includegraphics[width=0.40\textwidth]{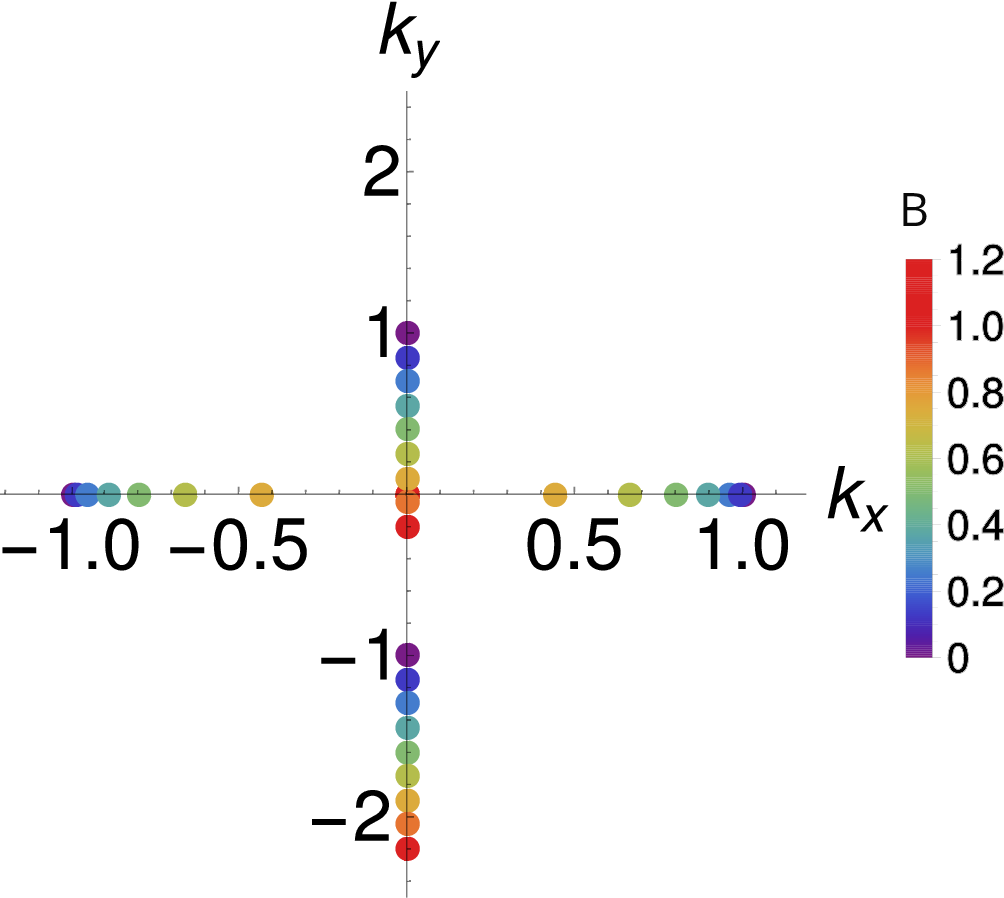}
\caption{\textbf{Position of EPs with magnetic field.} At $B=0$ the four EPs are located symmetrically about the origin (blue circles). With increasing $B$ two of them merge along $k_x$ and annihilate for $B=1$. The other two move along the negative $k_y$ direction. Here we set $t=1$, $t_{j1}=1$, $\lambda=1$ and $\gamma=1$.} 
\label{Fig_EP_B}
\end{figure}

\section{Magnetic field tuning of exceptional points}

We next propose and show that the positions of the EPs can be controlled by applying an external magnetic field, $B$, along the interface plane. We find that $B$ can not only change the EP positions asymmetrically, but also merge them beyond a critical value. For illustration, we include the external magnetic field, in our $d$-wave AM-FM junction, along $x$ direction, as shown in Fig.~\ref{Fig_junction}. Then, the Hamiltonian in Eq.~\ref{eq_d_j1} is augmented by a term of the form $B\sigma_x$. The conditions for EPs now read,

\begin{equation}
    \gamma^2=\lambda^2 k_x^2+(B+\lambda k_y)^2+4t_{j1}k_x^2k_y^2, \quad \textcolor{black}{t_{j1}} \gamma k_xk_y=0.
 \label{eq_dj1_B_ep}
\end{equation}

We find that, in this case, four EPs occur at $(0,\frac{\gamma-B}{\lambda})$, $(0,\frac{-\gamma-B}{\lambda})$, $(\sqrt{\frac{\gamma^2 -B^2}{\lambda^2}},0)$, $(-\sqrt{\frac{\gamma^2 -B^2}{\lambda^2}},0)$. Therefore, the coordinates of the EPs depend on $B$. The real and imaginary parts of the energy eigenvalues are plotted with increasing $B$ in Fig.~\ref{Fig_d_B1}(a) and (b). The corresponding phase rigidity, $r$, is shown in Fig.~\ref{Fig_d_B1}(c) and (d), along $k_x$ and $k_y$, respectively. We note that with increasing $B$, two of the EPs with opposite chirality move towards the origin and they collapse at the origin for a critical value of $B=\gamma$ and annihilate. The remaining two EPs move along the negative $k_y$ direction even beyond this critical value. For $B\geq \gamma$ the system is left with only two EPs. The motion of EPs in the $k_x-k_y$ plane with increasing $B$ is summarized in Fig.~\ref{Fig_EP_B}. We note that all the four EPs are located symmetrically about the origin for $B=0$. With increasing $B$, two EPs move nonlinearly along the $k_x$ axis and merge, while the other two move linearly in the same direction.

We observe that the position of the EPs can not only be tuned along $k_x$ and $k_y$ (using $B$), but also can be rotated about the origin by choosing different orientations of the AM Fermi surface with respect to the FM lead. For instance, a $\pi/4$ rotation is achieved with the AM Fermi surface shown in Fig.~\ref{Fig_junction}(b). This can be obtained by choosing different AM-specific spin-dependent hopping terms in the Hamiltonian \textcolor{black}{(for details see Appendix B).}
\textcolor{black}{We, furthermore, discovered that the positions of the emergent EPs can be adjusted to any arbitrary angle. We examine the generalized $d$-wave AM-FM junction~\cite{sun2023spin}, characterized by the effective NH Hamiltonian,}

\textcolor{black}{\begin{equation}
 \begin{aligned}
        \Tilde{H} =&H_\text{AM}^d + \Sigma_L(0)\\
              &= t(k_x^2 + k_y^2) \sigma_0 + 2t_{j1}k_xk_y\sigma_z
              +2t_{j2} (k_x^2- k_y^2) \sigma_z\\
              &+\lambda(k_y \sigma_x - k_x \sigma_y) + \Sigma_L.        
 \end{aligned}
 \label{eq_dj1j2_NH}
\end{equation}}

\textcolor{black}{Note that here terms proportional to both $t_{j1}$ and $t_{j2}$ are included in the Hamiltonian. For this case the conditions for obtaining EPs are given by,}

\textcolor{black}{
\begin{equation}
    \begin{aligned}
        \gamma^2=\lambda^2(k_x^2+k_y^2), \quad
         \gamma (t_{tj1}k_xk_y+t_{j2}(k_x^2-k_y^2))=0.
    \end{aligned}
    \label{eq_dj1j2_ep1}
\end{equation}}

\textcolor{black}{From Eq.~\ref{eq_dj1j2_ep1}, we find that the rotation angle, $\theta=\tan^{-1}(\frac{k_y}{k_x})$, depends on AM-specific spin-dependent hopping parameters as follows,}
\textcolor{black}{
\begin{equation}
    \tan\theta=\frac{-t_{j1}\pm\sqrt{t_{j1}^2+4t_{j2}^2}}{-2t_{j2}}.
    \label{eq_t2t1_reln}
\end{equation}}

\textcolor{black}{By a judicious choice of $t_{j1}$ and $t_{j2}$, we can obtain the EPs along a suitable angle in the $k_x-k_y$ plane.}

\begin{figure}
\centering
\includegraphics[width=0.48\textwidth]{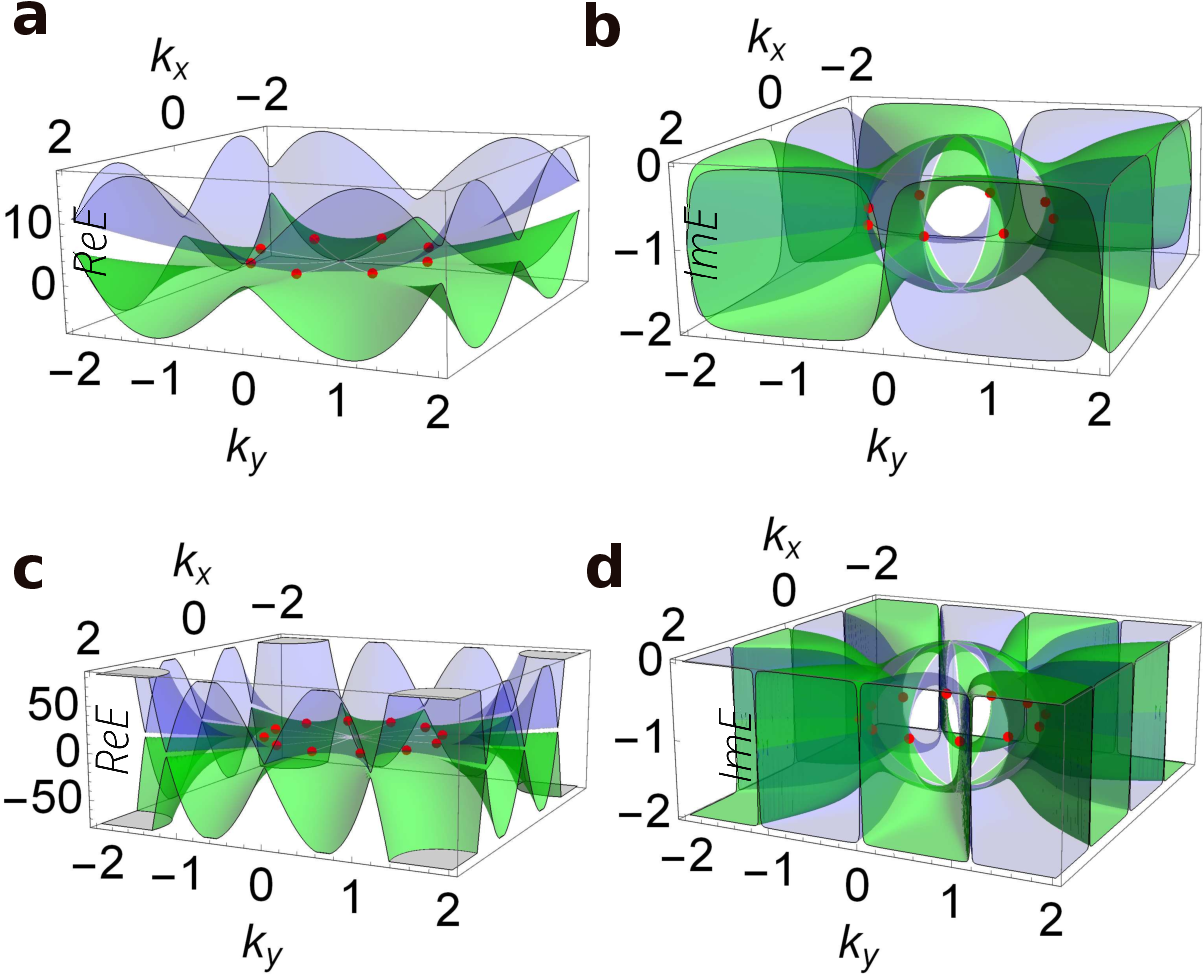}
\caption{\textbf{EPs in $g$- and $i$-wave AMs.} (a) Real and (b) imaginary parts of the energy for $g$-wave AM-FM junction. We obtain eight EPs. (c) Real and (d) imaginary parts of the energy for $i$-wave AM-FM junction. A total of twelve EPs emerge in this case. In general, the number of emerging EPs is twice the spin group integer of the AM. Here we choose $t=1$, $t_{j}=1$, $\lambda=1$ and $\gamma=1$.}
\label{Fig_g_i_B0}
\end{figure}

\section{Exceptional points in $g$- and $i$-wave altermagnets}

We further extend our analysis to AMs with higher harmonic spin-split Fermi surfaces. Let us first consider a $g$-wave AM-FM junction. The Hamiltonian for a $g$-wave AM reads~\cite{vsmejkal2022beyond}

\begin{equation}
\begin{aligned}
H_\text{AM}^g=t(k_x^2+k_y^2)\sigma_0+ 2t_{j}k_xk_y(k_x^2-k_y^2)\sigma_z
+\lambda(k_y\sigma_x-k_x\sigma_y).
\end{aligned}
\label{g_ham}
\end{equation}

Following our earlier analysis, as for the $d$-wave case, we obtain the conditions for EPs to be

\begin{equation}
    \gamma^2=\lambda^2(k_x^2+k_y^2)+4t_{j}^2k_x^2k_y^2(k_x^2-k_y^2)^2, \quad  \gamma t_{j}k_xk_y(k_x^2-k_y^2)=0.
    \label{eq_g_ep1}
\end{equation}

Strikingly distinct from the $d$-wave case, we find that a total of eight EPs emerge at such a junction. The location of EPs are given by $(0,\pm\frac{\gamma}{\lambda})$, $(\pm\frac{\gamma}{\lambda},0)$, $(\pm\frac{\gamma}{\sqrt{2}\lambda},\pm\frac{\gamma}{\sqrt{2}\lambda})$. These are shown by red dots in Fig.~\ref{Fig_g_i_B0}(a) and (b). We can extend this analysis to $i$-wave AMs \textcolor{black}{(see Appendix C for details of the Hamiltonian and location of EPs).}
In this case we find twelve EPs emerging at the AM-FM junction. The real and imaginary parts of the energy eigenvalues are presented in Fig.~\ref{Fig_g_i_B0}(c) and (d), respectively. \textcolor{black}{ We note that on setting $t_{j1} = 0$, the $g$-wave case reduces to the $d$-wave case. This is true for $i$-wave scenario as well. In fact, a non-zero hopping term $t_{j1}$ is essential to account for the effect of altermagnet symmetries on the occurrence of EP locations, which results in different number of EPs for different kinds of altermagnets. Therefore, this term is essential to capture the correct physics of the junction.} We observe that, in general, there appear twice the EPs as the spin group integer of the AM~\cite{vsmejkal2022emerging}. \textcolor{black}{ In case of bulk altermagnets, augmented by imaginary term proportinal to $\sigma_z$, we find that exceptional lines emerge. Their number is twice the number of spin group integer of the altermagnet. We note that the emergent EPs also follow the symmetry of the considered altermagnet.} We further note that an applied magnetic field should allow us to tune the EPs in these $g$- and $i$-wave AM-FM junctions, analogous to the $d$-wave case. 

\section{Summary and outlook}

We have proposed and demonstrated the emergence of NH physics, particularly EPs, in AM-FM junctions. We have revealed the presence of EPs in $d$-wave AM-FM junctions and shown their tunability using an external magnetic field. Our findings reveal that EPs can annihilate at a critical value of $B$. Additionally, we have found the tunability of EPs by varying the orientation of the AM Fermi surfaces. We discovered that the number of emergent EPs depends on the spin group integer of the underlying AM. Here we have focused on the EPs in AM-based NH systems.
\textcolor{black}{We note that our phase rigidity predictions should be clearly testable, as has been previously measured in several other systems~\cite{ding2016emergence,ding2022non}. Furthermore, other features of exceptional points, such as their distinctive energy scaling and concomitant Berry phases~\cite{dembowski2001experimental,dembowski2004encircling,tang2021direct}, would also be present in our proposed platform and could be directly measured in experiments. Furthermore, the non-Hermitian spectra in altermagnet-ferromagnet junctions may be directly amenable to surface spectroscopy measurements, such as angle resolved photoemission spectroscopy.}
In future, it could be worth exploring other NH aspects of altermagnetic systems, including NH skin effects, NH topology, and interplay with various types of dissipation mechanisms.

\section*{Acknowledgments}

We thank A. Banerjee, A. Bose, and S. Debadatta for useful discussions. M.A.R. is supported by a graduate fellowship of the Indian Institute of Science. A.N. acknowledges support from the DST MATRICS grant (MTR/2023/000021).

\bibliography{ref.bib}
\vspace{0.5cm}

\renewcommand{\theequation}{A\arabic{equation}}
\renewcommand{\thesection}{A\arabic{section}}
\setcounter{equation}{0}

\section*{Appendix A:Exceptional points in tight binding model of $d$-wave altermagnets}
\textcolor{black}{Here we consider the full tight-binding model for the altermagnet with the self energy arising from the ferromagnetic lead. The full tight binding model with self energy term is given by,}

\textcolor{black}{
\begin{equation}
\begin{aligned}
\Tilde{H} = & t(\cos^2 k_x + \cos^2 k_y) \sigma_0 + 2t_{j1} 
               \sin k_x \sin k_y \sigma_z\\
            &+\lambda(\sin k_y \sigma_x - \sin k_x \sigma_y) -i\gamma\sigma_z.        
 \label{eq_Full_TB}    
\end{aligned}
\end{equation}}
\textcolor{black}{The corresponding EP conditions are obtained to be,}
\textcolor{black}{
\begin{equation}
\begin{aligned}
    &\gamma^2=\lambda^2(\sin^2 k_x+ \sin^2 k_y)+4t_{j1}^2 \sin^2 k_x \sin^2 k_y, \\
    &\gamma t_{j1} \sin k_x \sin k_y=0,    
\end{aligned}
 \label{eq_EP_Full_TB}
\end{equation}}

\textcolor{black}{which give the EP locations at $(0,\pm \sin^{-1}(\gamma/\lambda))$, $(\pm \sin^{-1}(\gamma/\lambda),0)$. Therefore, our results are indeed valid considering the full tight-binding model.}

\begin{figure}[t]
\centering
\includegraphics[width=0.48\textwidth]{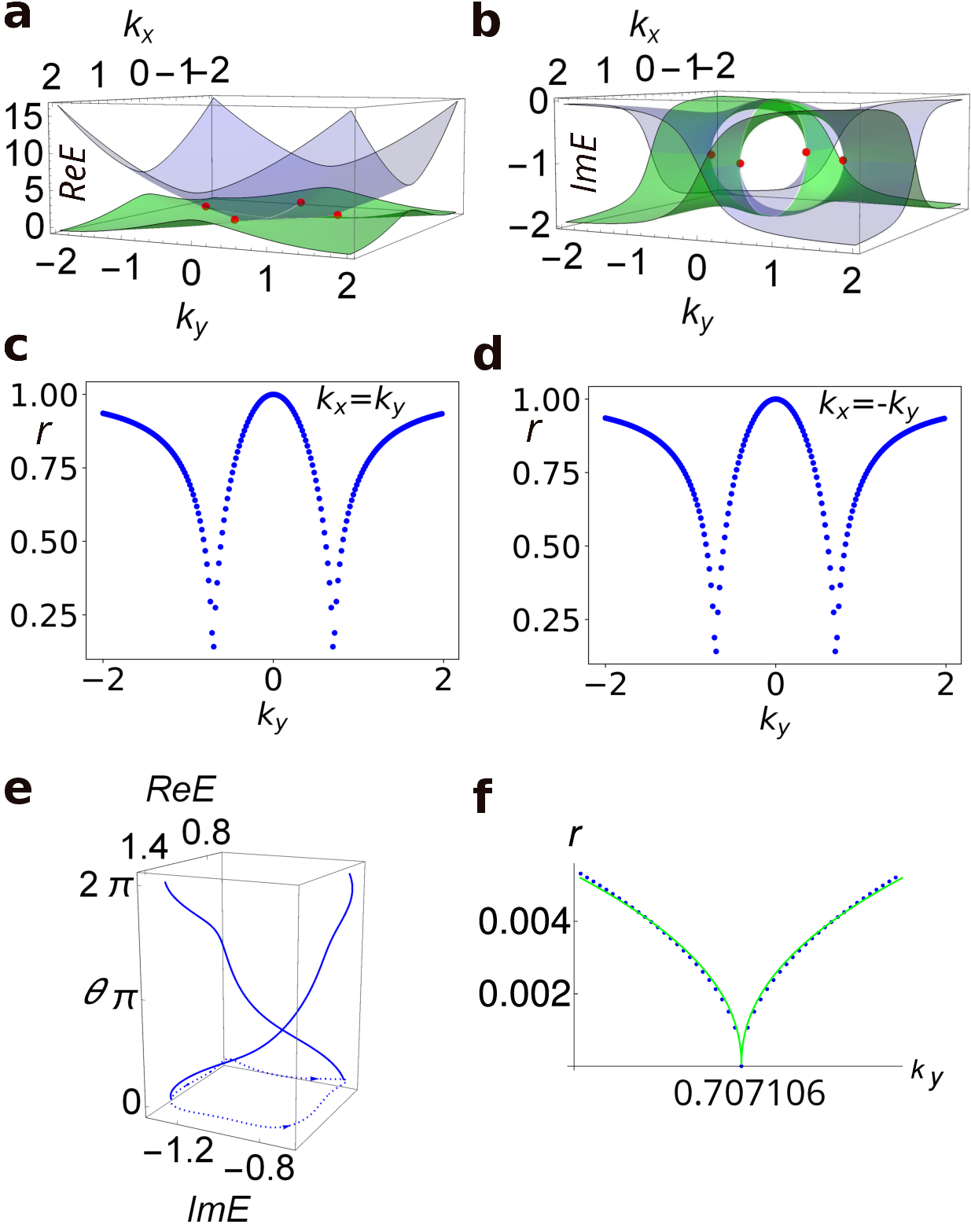}
\caption{\textbf{EPs in $d$-wave AM-FM junction with $\pi/4$ rotation.} (a) Real and (b) imaginary parts of the energy eigenvalues. At the red dots both the eigenvalues merge. Phase rigidity, $r$, along (c) $k_x$ and (d) $k_y$, respectively. Note that it goes to zero at ($\frac{1}{\sqrt{2}},\pm\frac{1}{\sqrt{2}})$, ($\pm\frac{1}{\sqrt{2}},\frac{1}{\sqrt{2}})$ and reaches one away from these coordinates. This indicates the emergence of EPs. We note that the EPs lie along $k_x=k_y$ and $k_x=-k_y$, i.e., they are rotated by an angle of $\pi/4$ compared to the $d$-wave case presented in the main text. (e) Vorticity around the EP at $(\frac{1}{\sqrt{2}},\frac{1}{\sqrt{2}})$. The braiding of the bands confirms the emergence of the EPs. (f) Scaling of phase rigidity around the EP. Blue dots are the calculated values and the solid green line represents the fitted function. From the fitted function we obtain an exponent of $\sim$0.46, close to that expected for a second-order EP. We choose $t=1$, $t_{j2}=1$, $\lambda=1$ and $\gamma=1$.}
\label{Fig_d_j2_B0}
\end{figure}

\section*{Appendix B: Tuning exceptional points in $d$-wave altermagnets}

We can tune the positions of EPs in $k$ space by altering the Fermi surface of the AM as shown in Fig. 1(b) of the main text, which is rotated by an angle of $\pi/4$ with respect to the Fermi surface shown in Fig. 1(a) of the main text. This rotates the EP positions by an angle of $\pi/4$. This can be achieved by choosing a different $d$-wave AM-FM junction effective NH Hamiltonian, which reads as follows,

\begin{equation}
\begin{aligned}
\Tilde{H} = &H_\text{AM}^d + \Sigma_L(0)\\
         &= t(k_x^2 + k_y^2) \sigma_0 + 2t_{j2} (k_x^2- k_y^2) \sigma_z \\
         &+\lambda(k_y \sigma_x - k_x \sigma_y) + \Sigma_L.        
 \label{S:eq_dj2_NH}    
\end{aligned}
\end{equation}

Here $t_{j2}$ is AM-specific spin-dependent hopping parameter distinct from $t_{j1}$. The other terms in the Hamiltonian are as already defined in the main text. Following similar calculations as in the main text, the conditions for the emergence of EPs turn out to be,

\begin{equation}
        \gamma^2=\lambda^2(k_x^2+k_y^2)+4t_{j2}^2(k_x^2-k_y^2)^2,
     \quad
        \gamma t_{j2}(k_x^2-k_y^2)=0.
    \label{eq_dj2_ep1}
\end{equation}

From Eq.~\ref{eq_dj2_ep1}, we find that the four EP positions are ($\frac{\gamma}{\sqrt{2}\lambda}$,$\pm \frac{\gamma}{\sqrt{2}\lambda}$) and ($ \pm \frac{\gamma}{\sqrt{2}\lambda}$,$\frac{\gamma}{\sqrt{2}\lambda}$). So, the EP positions not only come closer to the origin, but are also rotated by an angle of $\pi/4$, as compared to the $d$-wave AM whose Fermi surface orientation is as shown in panel Fig. 1(a) of the main text. We set $t=1$, $t_{j2}=1$, $\lambda=1$ and $\gamma=1$, which leads to four EPs at ($\frac{1}{\sqrt{2}},\pm\frac{1}{\sqrt{2}})$, ($\pm\frac{1}{\sqrt{2}},\frac{1}{\sqrt{2}})$ in the $k_x-k_y$ plane. 

We further confirm the emergence of EPs by plotting the real and imaginary parts of the energy eigenvalues in Fig.~\ref{Fig_d_j2_B0}(a) and (b), respectively. We observe that at the red dots, the eigenvalues merge. We further investigate the phase rigidity $r$ in Fig.~\ref{Fig_d_j2_B0}(c) and (d) along $k_x$ and $k_y$, respectively. We note that $r$ takes the value of zero at the positions of the red dots (EPs), as expected. In Fig.~\ref{Fig_d_j2_B0}(e), we plot the vorticity around the EP and the interchange of bands confirms the emergence of the EP. We further investigate the scaling of the phase rigidity near the EP. In Fig.~\ref{Fig_d_j2_B0}(f) the blue dots show the calculated $r$ values and the solid green line is the fitted function. The fitted function yields the exponent to be $\sim$0.46, which is close to the exponent expected for a second-order EP~\cite{bulgakov2006phase,rotter2009non}.

\section*{Appendix C: Exceptional points in $i$-wave altermagnets}

We consider an $i$-wave AM~\cite{vsmejkal2022beyond}. The corresponding AM-FM junction is described by the following NH Hamiltonian,

\begin{equation}
\begin{aligned}
H_\text{AM}^i= &t(k_x^2+k_y^2)\sigma_0+ 2t_{j}k_xk_y(3k_x^2 - k_y^2)(3k_y^2- k_x^2)\sigma_z \\
        &+ \lambda(k_y\sigma_x-k_x\sigma_y)+ \Sigma_L.
\end{aligned}
\end{equation}

Following similar calculations as the $d$- and $g$-wave cases, we find the following conditions for the emergence of EPs,

\begin{equation}
\begin{aligned}
&\gamma^2=\lambda^2(k_x^2+k_y^2)+4t_{j}^2k_x^2k_y^2(3k_x^2-k_y^2)^2(3k_y^2-k_x^2)^2,
\\
&\gamma t_{j}k_xk_y(3k_x^2-k_y^2)(3k_y^2-3k_x^2)=0.    
\end{aligned}        
\label{eq_i_ep1}
\end{equation}

From Eq.~\ref{eq_i_ep1}, we obtain the location of the EPs as $(0,\pm\frac{\gamma}{\lambda})$, $(\pm\frac{\gamma}{\lambda},0)$, $(\pm\frac{\gamma}{2\lambda},\pm\frac{\sqrt{3}\gamma}{2\lambda})$, $(\pm\frac{\sqrt{3}\gamma}{2\lambda},\pm\frac{\gamma}{2\lambda})$. We note that a total of twelve EPs emerge for the $i$ wave AM-FM junction. These are indicated by red points in Fig. 5(c) and (d) of the main text.

\end{document}